\documentclass[twocolumn,english,aps,prl,showpacs,longbibliography,showkeys,notitlepage,superscriptaddress]{revtex4-1}
\usepackage[T1]{fontenc}
\usepackage{lmodern}
\usepackage{color}
\usepackage{babel}
\usepackage{amsmath}
\usepackage{amssymb}
\usepackage{graphicx}

\RequirePackage[colorlinks=true,linkcolor=blue]{hyperref}
\usepackage{dsfont}
\usepackage{cleveref}
\usepackage{natbib}
\usepackage{bm}
\usepackage{blindtext}
\usepackage{pdfpages}

\usepackage{amsfonts}
\usepackage{babel}
\usepackage{braket}
\usepackage{verbatim}
\usepackage{amstext}
  
\makeatletter
\AtBeginDocument{\let\LS@rot\@undefined}
\makeatother

\begin{document}

\title{Optimized cavity-mediated dispersive two-qubit gates between spin qubits}

\author{M. Benito}
\address{Department of Physics, University of Konstanz, D-78457 Konstanz, Germany}
\author{J. R. Petta}
\address{Department of Physics, Princeton University, Princeton, New Jersey 08544, USA}
\author{Guido Burkard}
\address{Department of Physics, University of Konstanz, D-78457
  Konstanz, Germany}


\begin{abstract}
The recent realization of a coherent interface between a single
electron in a silicon quantum dot and a single photon trapped in a
superconducting cavity  opens the way for implementing photon-mediated
two-qubit entangling gates.
In order to couple a spin to the cavity electric field some type of spin-charge hybridization is needed, 
which impacts  spin control  and coherence. 
In this work we propose a cavity-mediated two-qubit gate and calculate cavity-mediated entangling gate fidelities in the dispersive regime,
accounting for errors due to the spin-charge hybridization, as well as
photon- and phonon-induced decays. 
By optimizing the degree of spin-charge hybridization, we show that
two-qubit gates mediated by cavity photons are capable of
reaching fidelities exceeding $90\%$ in present-day device
architectures. High iSWAP gate fidelities are
achievable even in the presence of charge noise at the level of $2\,\mu\text{eV}$.
\end{abstract}

\maketitle

\textit{Introduction.}
Recent advances in semiconductor fabrication, manipulation and readout
techniques have situated  spin qubits among the most promising
candidates for quantum information
processing~\cite{Hanson2007,Kloeffel2013,Awschalom2013}. The degree of control over 
single-electron spin qubits and their exchange interaction has allowed  high fidelity single \cite{Veldhorst2014,Takeda2016,Yoneda2018} and two-qubit \cite{Veldhorst2015, Zajac2017,Watson2018,Huang2018,Xue2018}  
gates.  Moreover,  recent improvements in the fabrication of
semiconductor quantum dots  (QDs) has pushed  the limits of QD modules
to sizable linear \cite{Zajac2016,Volk2019} and two-dimensional
\cite{Uditendu2018,Mortemousque2018}   arrays, which could allow not
only the implementation of multielectron encoded qubits  but also
intra-module operations and electron transfer \cite{Mills2018}. 
Several proposals exist to create a modular quantum architecture \cite{Taylor2005,Kimble2008} with all-to-all connectivity, which permits universal distributed   quantum computation  and high tolerances in error-correcting codes~\cite{Preskill1998}.

Within the approach of circuit quantum electrodynamics (QED)~\cite{Blais2004,Wallraff2004,Childress2004,2017arXiv170900466C,Delbecq2011}, the microwave field of a superconducting transmission line resonator (or cavity) mediates interactions between qubits separated by macroscopic distances, allowing a fully scalable and modular quantum information processing device~\cite{DiCarlo2009}. 
Although  electron spin qubits in semiconductor QDs promise long coherence times and potential for scalability, these photon-mediated interactions have not yet been  demonstrated for spin qubits.

In this paper, we theoretically describe
photon-mediated coupling of spin qubits and
calculate the achievable two-qubit gate fidelities.
Due to the small magnetic dipole of a single electron spin, some degree of spin-charge hybridization is needed to achieve a sizable coupling to the electric field of the cavity, a step which has recently been  demonstrated for single-electron spin qubits in a double quantum dot (DQD)~\cite{Viennot2015,Mi2018,Samkharadze2018} and for   a three-electron spin qubit in a triple QD~\cite{Landig2018}.
Here we focus on the single-electron Loss-DiVincenzo
 qubit~\cite{Loss1998}, where the mixing of orbital motion and spin is induced by an externally imposed magnetic field gradient,  and show that the spin qubit outperforms the intrinsic charge qubit both in the resonant and dispersive regime for levels of decoherence encountered in state-of-the-art devices. 

Although the spin-photon coupling strength is only a fraction of the charge-photon coupling~\cite{Mi2018,Samkharadze2018}, the spin decoherence is much slower and can be made comparable to the cavity loss rate, which allows for the optimization of the resonant coupling~\cite{Beaudoin2016,Benito2017}.
In the dispersive regime the qubit transition frequency is detuned
with respect to the photon frequency. In this case we find that the
externally-controllable spin-charge hybridization allows for optimal
detuning values implying high-fidelity  two-qubit gates. The
dispersive coupling scheme demands a relatively small degree of
spin-charge hybridization and will benefit enormously from the use of isotopically purified $^{28}\text{Si}$ material~\cite{Tyryshkin2012,Zwanenburg2013,Veldhorst2014,Yoneda2018}.

\begin{figure}
\includegraphics[width=1\columnwidth]{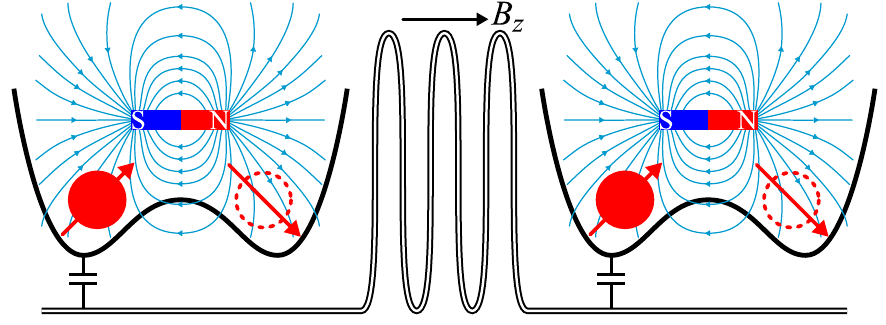}

\protect\caption{\label{fig:figure1} 
Schematic illustration of  two DQDs, represented as  double-well
potentials,
capacitively coupled to the same microwave superconducting resonator
(cavity) and influenced by a homogeneous external magnetic field
$B_z$.
The micromagnet has a field component that adds to $B_z$ and a transverse field gradient $b_x$.
The DQDs are tuned into the symmetric single-electron regime, with 
one electron delocalized across each DQD. 
}
\end{figure}

\textit{Model.}
We consider two DQDs capacitively coupled to the same cavity mode with  frequency $\omega_c$ (Fig.~\ref{fig:figure1}).  
Both DQDs are electrically tuned into the symmetric single-electron regime, with the electronic charge  distributed between the two QDs. The 
 tunnel coupling $t_c$ and the energy level detuning  $\epsilon$ can be electrically controlled.
 The electron  is capacitively coupled to the cavity electric field~\cite{Childress2004,Petersson2012,Bruhat2018,Stockklauser2017,Mi2017b}.
An  externally applied global magnetic field $B_z$  Zeeman splits the spin states and magnetizes the  micromagnets located near the DQDs.
The longitudinal component of the micromagnet field adds to $B_z$ and the micromagnet generates a transverse field gradient $b_x$ (typically  of order $\lesssim 1\,\text{mT/nm}$).
Each of the  DQDs can be described with a model 
Hamiltonian $\tilde{H}=\tilde{H}_{0}+\tilde{H}_{I}$ with
\begin{equation}
  \tilde{H}_{0}=t_c\tilde{\tau}_x +\frac{\epsilon}{2}\tilde{\tau}_z
  +\frac{B_z}{2}\tilde{\sigma}_z+\frac{b_x}{2}\tilde{\sigma}_x\tilde{\tau}_z+\omega_c a^{\dagger}a\ ,\label{eq:H0-eps0} 
\end{equation}
where $\tilde{\tau}_k$ and $\tilde{\sigma}_k$, for $k=x,y,z$, are the position and spin  Pauli matrices, respectively,
and  $a,a^{\dagger}$ are the cavity photon operators.
In the following, we will study the case of a symmetric DQD with $\epsilon=0$,
unless noted otherwise.
Here $\hbar=1$ and the magnetic fields are given in energy units. 
The light-matter interaction is described by $\tilde{H}_{I} = g_c\tilde{\tau}_z (a+a^{\dagger})$,
 where $g_c$ is the electric dipole coupling strength between the DQD
 electron and  a cavity  photon.
 Due to the spin-orbit effect induced by $b_x$, the electron spin dynamics are  sensitive to the cavity electric field~\cite{Cottet2010,Hu2012,Beaudoin2016,Benito2017}. 
 
\begin{figure}
\includegraphics[width=1\columnwidth]{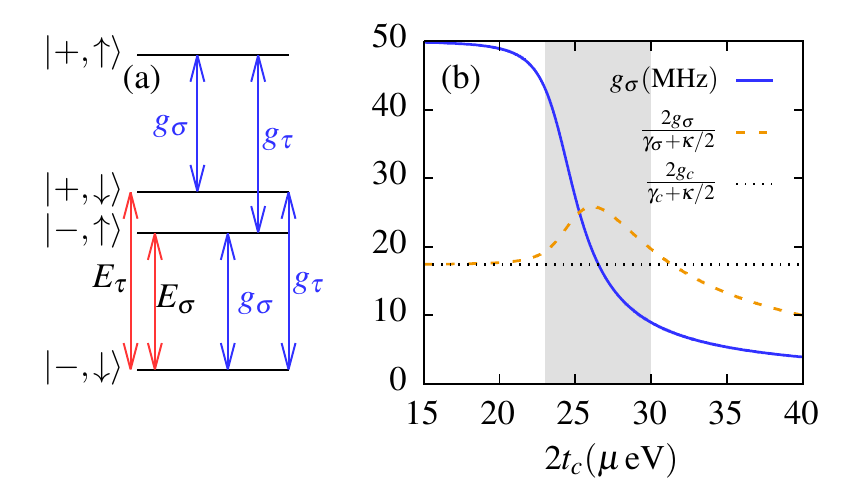}

\protect\caption{\label{fig:newfigure2} 
(a) Energy levels, $\ket{\pm,\uparrow(\downarrow)}=\ket{\pm}\otimes\ket{\uparrow(\downarrow)}$, with $\tau_z\ket{\pm}=\pm\ket{\pm}$ and $\sigma_z\ket{\uparrow(\downarrow)}=\pm\ket{\uparrow(\downarrow)}$,  for an electron localized in a DQD under the influence of an inhomogeneous magnetic field. The energy difference (red arrow) between ground and first (second) excited states is $E_{\sigma}$ ($E_{\tau}$). The blue arrows represent coherent coupling via cavity photons, with different coupling strengths $g_{\sigma}$ and $g_{\tau}$. (b) Coupling strength $g_{\sigma}$ (continuous blue line) as a function of the tunnel splitting $2 t_c$ for a fixed magnetic field profile; $B_z=24\,\mu\text{eV}$ and $b_x=2\,\mu\text{eV}$. 
 Also shown are the ratio between the coupling and decoherence for the spin (dashed orange line) and charge (dotted black line) qubit, for the parameters  $g_c=50\,\text{MHz}$, $\gamma_c=5\,\text{MHz}$, and $\kappa=1.5\,\text{MHz}$.
The shaded area indicates the regime where the spin qubit outperforms the charge qubit.
}
\end{figure}
 
 In the following we work in the basis that diagonalizes $\tilde{H}_0$, where bonding and antibonding states of the DQD with opposite spin are hybridized~\cite{Benito2017}. We  define new Pauli matrices $\tau_k, \sigma_k$ in terms of which  the  transformed model Hamiltonian reads
\begin{align}
H_0&=\frac{E_{\tau}}{2}\tau_z+\frac{E_{\sigma}}{2}\sigma_z+\omega_c a^{\dagger}a\ ,\\
H_{I} &= \left(-g_{\tau}\tau_x+g_{\sigma}\sigma_x\tau_z\right) (a+a^{\dagger})\ , \label{eq:dipole}
\end{align}
with the energy levels at $E_{\tau(\sigma)}=r_+\pm r_-$, where  $r_{\pm}=\sqrt{(2t_c\pm B_z)^2+b_x^2}/2$, 
coupled to the cavity with strength
 $g_{\tau}=g_c \cos{\bar{\phi}} $ and  $g_{\sigma}=g_c
 \sin{\bar{\phi}} $, where $\bar{\phi}=\left(\phi_++\phi_-\right)/2$
 is the spin-charge mixing angle, with 
 $\phi_{\pm}=\arctan{\left[b_x/(2t_c\pm B_z)\right]}$.
The corresponding level scheme is shown in Fig.~\ref{fig:newfigure2}~(a). In Fig.~\ref{fig:newfigure2}~(b) we show how the coupling strength $g_{\sigma}$ decreases with increasing tunnel coupling $t_c$ for a given magnetic field profile, as a consequence of the increasing spin character of the qubit.  
If the qubit energy $E_{\sigma}$ equals the photon frequency $\omega_c$,
coherent state transfer between a cavity photon and the qubit is possible whenever the coupling strength $g_{\sigma}$
overcomes  the total decoherence rate $\gamma_{\sigma}+\kappa/2$.
Typical cavity photon frequencies are around $\omega_c/2\pi \sim 6\,{\rm GHz} \sim 25\,\mu{\rm eV}$.
 In Fig.~\ref{fig:newfigure2} (b), we also show the ratio between coupling and decoherence  for the system with (dashed line) and without  (dotted line) magnetic fields, i.e., for a spin and a charge qubit respectively, where we have assumed that the decoherence rate of the former is inherited from the hybridization with charge, therefore $\gamma_{\sigma}=\gamma_c \sin^2{\bar{\phi}}$, where $\gamma_c$ is the total charge qubit decoherence rate. 
 Although the charge qubit can be made sufficiently coherent to reach
 the strong coupling regime
 \cite{Mi2017b,Stockklauser2017,Woerkom2018}, the spin qubit
 overcomes its performance in the shaded gray area in
 Fig.~\ref{fig:newfigure2} (b) ($2t_c\gtrsim B_z$) if
 $\gamma_c>\kappa/2$, where  $\kappa$ is the cavity loss rate.
 More precisely, we find that the spin qubit performs better than the charge qubit when
 $\sin{\bar{\phi}}>\kappa/2\gamma_c$ in a finite  interval around $2t_c \approx B_z$
 where $\bar{\phi}$ denotes the spin-charge mixing angle defined
 above. In this regime, the gain from
 using the spin with a long coherence time overcompensates the
 decrease in coherence from spin-charge hybridization.
 Therefore, the advantage to be gained from using spin rather
   than charge as a qubit is twofold: (i) In the regime
   indicated by the shaded region in Fig.~\ref{fig:newfigure2} (b), the exchange of
   quantum information between the qubit and the cavity photons is
   more efficient for the spin, and (ii) the spin-charge and
   spin-photon couplings can be switched off efficiently by
   controlling $t_c$ and $\epsilon$, thus reaching a memory qubit
   regime where the spin qubit is far more coherent than the charge qubit.
 Importantly, in order to control the interaction times, there are two
 mechanisms to electrically switch off the spin-photon coupling: a)
 increasing the tunnel coupling,
 as shown in Fig.~\ref{fig:newfigure2},
 or b) by increasing the level detuning $\epsilon$ and thereby
   reducing the amount of charge admixture across the DQD,
   such that the charge-photon coupling is reduced  as $\tilde{g}_c=  2t_c g_c/\sqrt{\epsilon^2+4t_c^2}$.

When  two subsystems (denoted with index $i=1,2$) as described above
are coupled to the same cavity,
the cavity photons can  induce a long distance coupling between their spins. 
In the resonant regime ($E_{\sigma}=\omega_c$) there is a
collectively-enhanced two-qubit coupling that can be observed in a
transmission experiment~\cite{Majer2007,Fink2009,Woerkom2018}.
Here we
investigate the dispersive regime, where the photon frequency is
detuned from the qubit transition frequency $E_{\sigma}$ and a
coherent long-distant  interaction is mediated by the exchange of
virtual photons~\cite{Blais2004,Imamoglu1999,Burkard2006}. This
mechanism results in a smaller effective coupling but is
less sensitive to photon loss in the cavity.

\begin{figure*}
\includegraphics{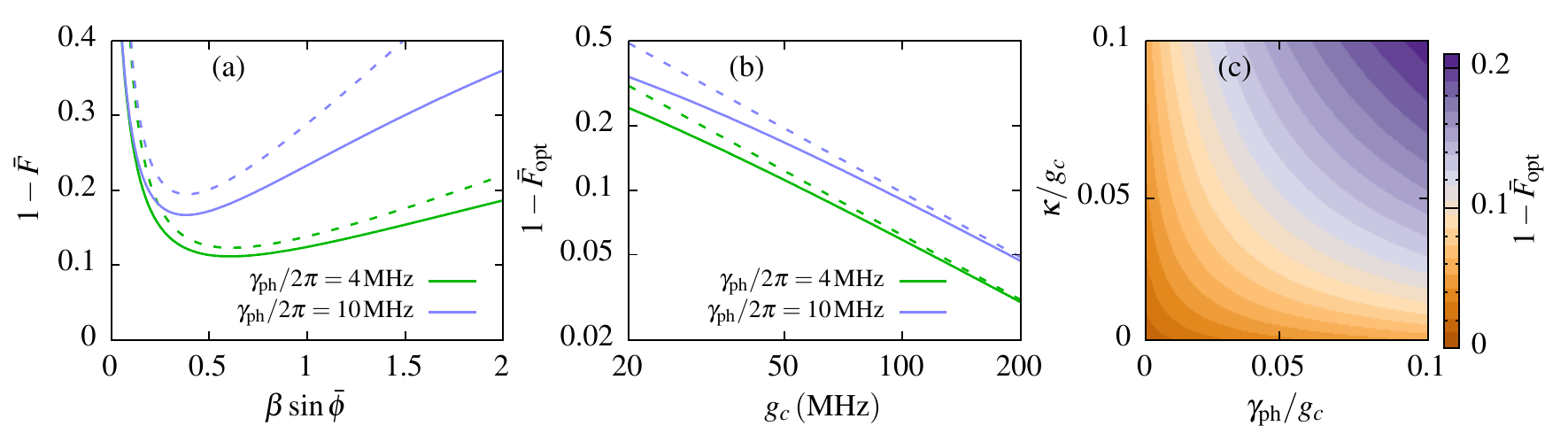}
\caption{\label{fig:figure3} 
(a) Average infidelity of an iSWAP gate  as a function of  $\beta\sin{\bar{\phi}}$, where $\beta=\Delta/g_{\sigma}$ is  the qubit-photon detuning in units of the spin-photon coupling,  and $\bar{\phi}$ is the spin-charge mixing angle, for the indicated values of $\gamma_{\text{ph}}$. For this plot, we have chosen the charge coupling strength $g_c/2\pi=50\,\text{MHz}$. (b)  Double-logarithmic plot of the optimal average infidelity  $1-\bar{F}_{\text{opt}}$ as a function of $g_c$ for the same  values of $\gamma_{\text{ph}}$. In (a) and (b) the photon loss rate is $\kappa/2\pi=1.5\,\text{MHz}$ and the continuous  lines correspond to  Eq.~\eqref{eq:F} while the dashed lines correspond to the approximated result in Eq.~\eqref{eq:fidelity-approx}. (c) Optimal average infidelity  $1-\bar{F}_{\text{opt}}$ as a function of $\gamma_{\text{ph}}/g_c$ and $\kappa/g_c$.
}
\end{figure*}

\textit{Dispersive regime.}
The light-matter interaction Hamiltonian \eqref{eq:dipole} couples subspaces with different number of
cavity photons. If the coupling strengths are small such that
$g_{\tau(\sigma)}\ll |E_{\tau(\sigma)}-\omega_c|$, we can decouple
such subspaces to a desired order using a Schrieffer-Wolff
transformation \cite{SI}.
From now on we assume  $2t_c > B_z$, which ensures that the lower-energy subspace constitutes a qubit with  a good coherence  inherited from its spin character.
 Therefore the  most interesting operating regime is the one with   the cavity frequency $\omega_c$ being close to  $E_{\sigma}$, therefore fulfilling the condition, $g_{\tau}|\Delta|/|E_{\tau}-\omega_c|\ll  g_{\sigma}\ll |\Delta|$, with the detuning  $\Delta=E_{\sigma}-\omega_c$.

Assuming identical DQDs, we find to first order in the perturbative parameter $g_{\sigma}/\Delta$, 
 in the limit of an empty cavity and within a rotating-wave
 approximation (RWA) \cite{SI}, the following dispersive  Hamiltonian 
\begin{align}
 H^{\text{d}} &\simeq\frac{E_{\tau}}{2}\sum_{i=1,2}\tau_z^{(i)}+\frac{E_{\sigma}+g_{\sigma}^2/\Delta}{2}\sum_{i=1,2}\sigma_z^{(i)} \nonumber \\
  &+\frac{g_{\sigma}^2}{\Delta}\left(\sigma_+^{(1)}\sigma_-^{(2)}+\sigma_-^{(1)}\sigma_+^{(2)}\right)\tau_z^{(1)}\tau_z^{(2)} \ , \label{eq:Hd-z-terms}
\end{align}
where the Pauli  matrices  correspond to dressed states.
The goal now is to harness the spin-spin long-distance coupling term
to perform a two-qubit gate.
A  coupling Hamiltonian of the form $\sim g_{\sigma}^2\left(\sigma_+^{(1)}\sigma_-^{(2)}+h.c.\right)/\Delta$ performs an 
 iSWAP gate at gate times
 $t_g=(4m+1)\frac{\pi}{2}\frac{|\Delta|}{g_{\sigma}^2}$
 for any integer $m$, e.g. $\ket{\uparrow,\downarrow}\rightarrow i \ket{\downarrow,\uparrow}$.
 A  CNOT entangling gate can be constructed with two iSWAP gates and single qubit rotations~\cite{Imamoglu1999,Schuch2003}. 
 To estimate how well such a gate can perform we  take into account
 three sources of infidelity: (1) The full system Hamiltonian also
 contains  cavity-mediated long-distance $\tau$-coupling, and
 $\sigma^{(i)}$-$\tau^{(i)}$ coupling within a DQD as well as  between
 distant DQDs ($\sigma^{(1)}$-$\tau^{(2)}$) \cite{SI};
 (2) Even for a material system with very low magnetic noise,  spin-charge hybridization makes the spin qubits susceptible to  charge noise. Therefore, the electron-phonon interaction and other charge fluctuations commonly present in semiconductor nano-structures  will contribute to decoherence; (3) As the qubits are dressed by photonic excitations, the cavity damping will also contribute to  qubit decoherence.

In order to capture dissipative effects, we model the system consisting of two DQDs using a master equation in the dispersive regime~\cite{Boissonneault2009}.
 If we assume that the system is prepared in the lower energy charge sector, we can derive an effective equation for the partial density matrix $\rho^{\text{d}}_{\sigma}$, corresponding to the spin degrees of freedom,  $\sigma^{(1)}$ and $\sigma^{(2)}$. In the rotating frame the master equation can be written as
   \begin{align}
 \dot{\tilde{\rho}}^{\text{d}}_{\sigma}\simeq&-i\left[\frac{g_{\sigma}^2}{\Delta}\left(\sigma_+^{(1)}\sigma_-^{(2)}+\sigma_-^{(1)}\sigma_+^{(2)}\right),\tilde{\rho}^{\text{d}}_{\sigma}\right]\nonumber \\
& +\frac{\gamma_{\text{ph}}}{2} \sin^2\bar{\phi}\left({\cal{D}}\left[\sigma_-^{(1)}\right]+{\cal{D}}\left[\sigma_-^{(2)}\right]\right)\tilde{\rho}^{\text{d}}_{\sigma} \nonumber\\
&+\frac{\kappa}{2}\frac{g_{\sigma}^2}{\Delta^2}{\cal{D}}\left[\sigma_-^{(1)}+\sigma_-^{(2)}\right]\tilde{\rho}^{\text{d}}_{\sigma} \ ,  \label{eq:master-eq}
 \end{align}
 where $\mathcal{D}\left[c\right]$ represents the usual Lindblad superoperator $\mathcal{D}\left[c\right]\rho=2c\rho c^{\dagger}-c^{\dagger}c\rho-\rho c^{\dagger}c$.
The first term in Eq.~\eqref{eq:master-eq} describes the long-distance
coupling mediated by the cavity, with strength $g_{\sigma}^2/\Delta$,
while the second term corresponds to relaxation due to  phonon emission
with rate $\gamma_{\text{ph}}$ (at the qubit energy $E_{\sigma}$)  in
each DQD. Finally, the last term describes the Purcell relaxation,
i.e., relaxation of the qubits with rate $\kappa g_c^2/\Delta^2$ due to photon decay with  rate $\kappa$. 
Given the reported long   coherence  times of electron spins in silicon QDs,
we assume here that the spin qubit decoherence is mainly inherited
from the hybridization with charge and we neglect other decoherence
sources such as hyperfine interaction with nuclear spins \cite{Kloeffel2013}.

\textit{Results.}
With our effective  model describing the system dynamics, we can
estimate how accurately one  can expect to realize a two-qubit iSWAP
gate taking into account the amount of decoherence
encountered in present-day experiments. 
We introduce the average fidelity $\bar{F} = \overline{\langle \psi|
  \cal{E} [|\psi _{\rm i} \rangle]|\psi\rangle}$ as a measure of the quality of a quantum gate 
 which compares the targeted  pure state  $\ket{\psi}$ and the
 mixed state density matrix $\cal{E} [|\psi _{\rm i}
 \rangle]$ obtained from the gate $\cal{E}$ including decoherence,
 averaged over all possible pure input states $|\psi _{\rm i}\rangle$.
 To avoid a direct evaluation of the average over initial states, we
 use a method of calculating $\bar{F}$ via the fidelity of
 entanglement $F_e$,  and using the relation $\bar{F}=(4 F_e+1)/5$, valid for two-qubit gates~\cite{Nielsen2002,White2007,Elman2017}.

Choosing a qubit-cavity detuning $|\Delta|=\beta g_{\sigma}$ with
$\beta\gg 1$ to ensure the dispersive regime and a gate time corresponding to the shortest iSWAP gate, $t_g=\pi \beta/(2g_{\sigma})$, 
 we find for the average fidelity corresponding to Eq.~\eqref{eq:master-eq} 
   \begin{equation}
 \bar{F}=  \frac{1}{20}\left[4+\left(1+x\right)^2\left(1+x y\right)^2\right]
 \label{eq:F} 
 \end{equation}
 with 
 \begin{equation}
 x=e^{-\pi \gamma_{\text{ph}}\beta \sin{\bar{\phi}}/(4g_c)},\quad  y=e^{-\pi \kappa/(2\beta \sin{\bar{\phi}}g_c)},
 \end{equation}
 which for $\gamma_{\text{ph}}\beta \sin{\bar{\phi}},\kappa/(\beta \sin{\bar{\phi}}) \ll g_c$ can be approximated as   
\begin{equation}
 \bar{F}\simeq1-\frac{2\pi}{5 g_c}\left(\gamma_{\text{ph}}\beta  \sin\bar{\phi}+\frac{\kappa}{\beta \sin{\bar{\phi}}}\right) \ . \label{eq:fidelity-approx}  \end{equation}
This approximated result suggests that there is  
 an optimal value of $\beta \sin{\bar{\phi}}$, related to the detuning $\Delta$, that maximizes the average fidelity. 
This value is
   \begin{equation}
 ( \beta\sin\bar{\phi})_{\text{opt}}\simeq\sqrt{\frac{ \kappa  }{   \gamma_{\text{ph}}}} \ , \label{eq:sin-opt} \end{equation}
with a corresponding approximated average  fidelity of
   \begin{equation}
 \bar{F}_{\text{opt}}\simeq  1-\frac{4\pi}{5} \sqrt{\frac{\gamma_{\text{ph}}\kappa}{g_c^2}} \label{eq:Fopt}  \ .
 \end{equation}
%
 %
Although the optimal average  fidelity is determined by  charge qubit  and cavity parameters, via the cooperativity $C=g_c^2/\gamma_{\text{ph}}\kappa$, the role of the spin-charge hybridization is to enable the optimization in Eq.~\eqref{eq:sin-opt}, which is not accessible for  charge qubits ($\bar{\phi}= \pi/2$) unless $\gamma_{\text{ph}}\ll\kappa$ (because $\beta\gg 1$). 
%
%
%
From Eq.~\eqref{eq:sin-opt} we can extract that the spin qubit decoherence rate inherited from the charge at the optimal point is $\gamma_{\sigma}=\kappa/(2\beta^2)$.

In Fig.~\ref{fig:figure3}(a), we show the exact average infidelity,
Eq.~\eqref{eq:F}, compared with the approximation,
Eq.~\eqref{eq:fidelity-approx},  as a function of $\beta
\sin{\bar{\phi}}$ for a typical value of charge-cavity coupling
strength $g_c/2\pi=50\,\text{MHz}$ for different values of the
phonon-induced charge decoherence rate  $\gamma_{\text{ph}}$.
The best gate can be performed at the minimum of these curves,
 which can be found numerically for the exact expression  by demanding $\partial \bar{F}/\partial (\beta\sin{\bar{\phi}})=0$.
For currently available system parameters $g_c$, $\gamma_{\rm ph}$, and
$\kappa$, we find fidelities around $\bar{F}\approx 90\%$.
Improvements are possible via all three parameters.
E.g., typical values of $g_c$ are around $50\,{\rm MHz}$
\cite{Mi2018}, but can be increased beyond $100\,{\rm MHz}$ with the
use of high-impedance resonators (cavities) \cite{Stockklauser2017},
leading to $\bar{F}\approx 95\%$.  An even higher fidelity of
$\bar{F}\approx 99\%$
could be reached if e.g. $g_c\approx 250\,{\rm MHz}$ and
$\gamma_c\approx \kappa\approx 1\,{\rm MHz}$.
 In Fig.~\ref{fig:figure3}~(b)   we show the predicted average infidelity at the  optimal value $(\beta \sin{\bar{\phi}})_{\text{opt}}$ as a function of $g_c$ in a double-logarithmic plot. As expected,  the exact result coincides with the approximation, Eq.~\eqref{eq:Fopt},  for large $g_c$ and the average infidelity is inversely proportional to  $g_c$.
Finally, in Fig.~\ref{fig:figure3}~(c)  we show the exact predicted infidelity at the  optimal value $(\beta \sin{\bar{\phi}})_{\text{opt}}$ as a function of both $\gamma_{\text{ph}}/g_c$ and $\kappa/g_c$. 

\begin{figure}
\includegraphics{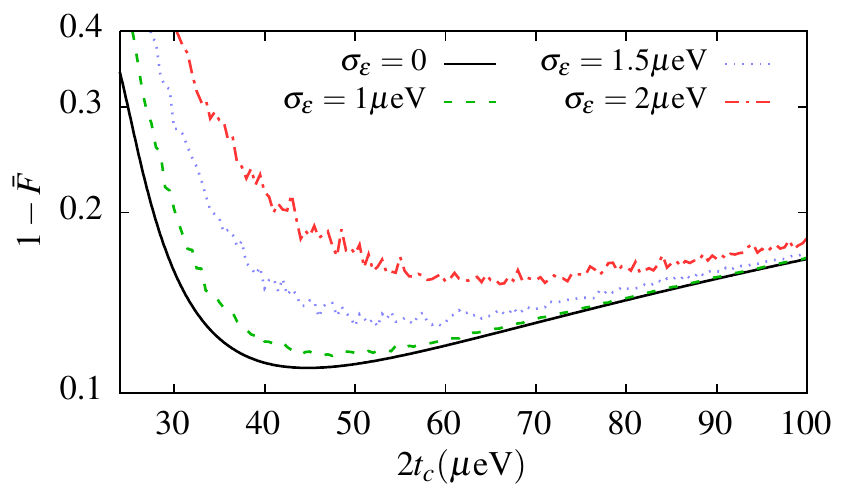}
\protect\caption{\label{fig:figure4} 
Average infidelity of an  iSWAP gate between spin qubits as a function of the tunnel splitting $2t_c$, with fixed   $b_x=2\,\mu\text{eV}$ and $\omega_c=24\,\mu\text{eV}$ and   in the dispersive regime, $\Delta=-10\, g_{\sigma}$. The charge-photon coupling is chosen as $g_c/2\pi=50\,\text{MHz}$, the phonon relaxation rate as  $\gamma_{\text{ph}}/2\pi=4\,\text{MHz}$ and the cavity loss rate as $\kappa/2\pi=1.5\,\text{MHz}$.   The continuous  line corresponds to  Eq.~\eqref{eq:F} while the other  lines correspond to the indicated values of low-frequency charge noise standard deviation $\sigma_{\epsilon}$.
}
\end{figure}

In current experiments, the magnetic field gradient $b_x$ and the cavity frequency $\omega_c$ are fixed, but it is possible to electrically tune the  tunnel coupling $t_c$ between the QDs, modifying in this way the spin-charge hybridization. 
Therefore, one can tune $t_c$ and the external magnetic field such that the optimal fidelity condition is fulfilled
and the spin qubits are in the dispersive regime, $ g_{\sigma}\ll |\Delta|$. 
 In Fig.~\ref{fig:figure4} (continuous line), we show the average
 infidelity $1-\bar{F}$ of the dispersive iSWAP gate as a function of
 $t_c$.
 The result for different values of $b_x$, $g_c$ and the comparison
 with the full master equation can be found in \cite{SI}.

\textit{Charge noise.}
The realistic entangling gates fidelities between spin qubits  are currently limited by fluctuations due to charge noise.
Since in our setup the qubits are at the ``sweet spot'' $\epsilon=0$, i.e., they are first-order insensitive to  onsite energies fluctuations (with amplitude $\delta_i$), the noise enters solely to second order.
Here we include a high-frequency charge noise contribution, which can be modeled by adding dephasing  Lindblad terms to the master equation~\eqref{eq:master-eq},
\begin{equation}
{\mathcal{L}}_{\phi}\rho_{\sigma}^{d}=\frac{\gamma_{\phi}}{4}\sin^2\bar{\phi} \left({\cal{D}}\left[\sigma_z^{(1)}\right]+{\cal{D}}\left[\sigma_z^{(2)}\right]\right)\rho_{\sigma}^{d}
\label{eq:highfnoise}\ ,
\end{equation} 
and a low-frequency component or quasistatic  noise that randomizes the qubit energies via  the Hamiltonian term
\begin{equation}
H_{\delta}=\frac{\delta_1^2 \sin^2{\bar{\phi}}}{4 E_{\sigma}}\sigma_z^{(1)}+\frac{\delta_2^2 \sin^2{\bar{\phi}}}{4 E_{\sigma}}\sigma_z^{(2)}  \label{eq:low-freq}\ .
\end{equation}
Accounting for high-frequency charge noise, the approximated result in Eq~\eqref{eq:fidelity-approx} needs to be revised  as $\gamma_{\text{ph}} \rightarrow \gamma_{\text{ph}}+\gamma_{\phi}$.
In order to calculate the  effect of the low-frequency noise, we
average the fidelity over a Gaussian distribution with standard
deviation $\sigma_{\epsilon}$ for the variables $\delta_i$,
with typical values
$\sigma_{\epsilon}\sim1-4\,\mu\text{eV}$~\cite{Petersson2010,Mi2018b}.
The mean value of Eq.~\eqref{eq:low-freq} only shifts the qubit energies and  is included  into the rotating frame transformation. 
 In Fig.~\ref{fig:figure4}, we show the average infidelity $1-\bar{F}$ of the dispersive iSWAP gate as a function of $t_c$
for  different levels $\sigma_{\epsilon}$ of low-frequency charge noise.

\textit{Conclusions.}
We have analyzed the performance of single-electron spin qubits in DQDs with respect to dispersive long-distance two-qubit gates mediated by virtual cavity photons. By solving a  model master equation, our results show that this implementation benefits from the spin-charge hybridization since this allows us to optimize the average iSWAP gate fidelity $\bar{F}$, even for the decoherence rates found in state-of-the-art experiments, where the qubit decoherence is worse than the photon decoherence. We  predict the degree of spin-charge hybridization, controlled via the tunnel coupling $t_c$ (Fig.~\ref{fig:figure4}), needed to optimize this gate, $\beta\sin{\bar{\phi}}\simeq\sqrt{\kappa/\gamma_{\text{ph}}}$, and explain how the spin qubit outperforms the DQD charge  qubit.

The analyzed setup is capable of reaching iSWAP gate fidelities exceeding $90\%$ with present-day device architectures.
We expect that the same kind of analysis can be readily applied to the triple QD spin-qubit strongly coupled to a resonator~\cite{Landig2018}.
The performance of other two-qubit
gates~\cite{Blais2007,Haack2010,Srinivasa2016} and other
qubit-resonator coupling schemes,
such as longitudinal coupling~\cite{Elman2017,Harvey2018,Bosco2019}, will be the subject of future studies.

\textit{Note added.}
While finalizing this work, we became aware of a recent related study
\cite{Warren2019} where the transitions to excited states due to the
influence of non-adiabatic effects during a cavity-mediated two-qubit
gate in the dissipationless (unitary) case were studied.

\begin{acknowledgments}
\textit{Acknowledgments.---}
This work has been supported by the Army Research Office grant W911NF-15-1-0149. 
\end{acknowledgments}

\bibliographystyle{apsrev4-1}
\bibliography{cavity-gate}



\section*{Supplementary material}

# Supplemental Material to "Optimized cavity-mediated dispersive two-qubit gates between spin qubits"

M. Benito,[1] J. R. Petta,[2] and Guido Burkard[1]

[1]*Department of Physics, University of Konstanz, D-78457 Konstanz, Germany*
[2]*Department of Physics, Princeton University, Princeton, New Jersey 08544, USA*

## I. HAMILTONIAN IN THE DISPERSIVE BASIS

To analyze the dispersive regime, we apply the Schrieffer-Wolff transformation $e^{-S}He^S$, with $S = S_+ a - S_- a^\dagger$ such that $[H_0, S] = -H_I$ to decouple the subspaces with different number of photons to first order in the perturbative parameters $g_{\tau(\sigma)}/|E_{\tau(\sigma)} - \omega_c|$. We assume an empty cavity such that $\langle a^\dagger a\rangle = 0$. The dispersive Hamiltonian then reads

$$H^{\rm d} \simeq H_0 + \frac{1}{2}\left[H_I, S\right] = \sum_{i=1,2} \left(\frac{\tilde{E}_\tau}{2}\tau_z^{(i)} + \frac{\tilde{E}_\sigma}{2}\sigma_z^{(i)} + S_x \sigma_x^{(i)}\tau_x^{(i)} + S_y \sigma_y^{(i)}\tau_y^{(i)}\right)$$
$$+ J_\tau \tau_x^{(1)}\tau_x^{(2)} + J_\sigma \sigma_x^{(1)}\tau_z^{(1)}\sigma_x^{(2)}\tau_z^{(2)} - J_{\tau,\sigma}\left(\sigma_x^{(1)}\tau_z^{(1)}\tau_x^{(2)} + \tau_x^{(1)}\sigma_x^{(2)}\tau_z^{(2)}\right), \quad (1)$$

with dressed-state energies $\tilde{E}_\alpha = E_\alpha + 2g_\alpha^2 \frac{E_\alpha}{E_\alpha^2 - \omega_c^2}$, spin-charge couplings $S_x = g_\tau g_\sigma \frac{E_\tau}{E_\tau^2 - \omega_c^2}$ and $S_y = g_\tau g_\sigma \frac{E_\sigma}{E_\sigma^2 - \omega_c^2}$ within the same DQD, and distant charge $J_\tau = \frac{2g_\tau^2 \omega_c}{E_\tau^2 - \omega_c^2}$, spin $J_\sigma = \frac{2g_\sigma^2 \omega_c}{E_\sigma^2 - \omega_c^2}$, and spin-charge $J_{\tau,\sigma} = \frac{g_\tau g_\sigma \omega_c (E_\sigma^2 + E_\tau^2 - 2\omega_c^2)}{(E_\tau^2 - \omega_c^2)(E_\sigma^2 - \omega_c^2)}$ couplings. The Pauli matrices $\tau_k, \sigma_k$ no longer correspond to the original qubits, but rather qubits that are dressed by excitations of the cavity.

As explained in the main text, the regime of interest to address the spin qubit is $\omega_c \approx E_\sigma$. In this regime, $g_\tau \Delta/|E_\tau - \omega_c| \ll g_\sigma \ll \Delta$ (with $\Delta = E_\sigma - \omega_c$), the Hamiltonian (1) to first order in $g_\sigma/\Delta$ reads

$$H^{\rm d} \simeq \sum_{i=1,2}\left(\frac{E_\tau}{2}\tau_z^{(i)} + \frac{E_\sigma + g_\sigma^2/\Delta}{2}\sigma_z^{(i)} + \frac{g_\sigma g_\tau}{2\Delta}\sigma_y^{(i)}\tau_y^{(i)}\right) + \frac{g_\sigma^2}{\Delta}\sigma_x^{(1)}\tau_z^{(1)}\sigma_x^{(2)}\tau_z^{(2)}. \quad (2)$$

Finally, we can neglect non-resonant terms using the rotating-wave approximation (RWA) and we obtain the Hamiltonian in Eq. (4) in the main text.

## II. CAVITY DAMPING

The cavity mode couples to an electromagnetic bath via the electric field operator $\propto a + a^\dagger$. If the bath is considered to be at zero temperature, it only induces relaxation on the cavity mode. Since the qubits are coupled to the cavity photons, they also suffer relaxation due to this bath. To find the dissipative terms we need to transform the electric field operator in the dispersive basis and trace out the bath. To first order in the perturbative parameters $g_{\tau(\sigma)}/|\omega_c - E_{\tau(\sigma)}|$ it reads

$$a + a^\dagger \to (a + a^\dagger) - \sum_{i=1,2}\left\{\tau_x^{(i)}\frac{2g_\tau \omega_c}{E_\tau^2 - \omega_c^2} - \sigma_x^{(i)}\tau_z^{(i)}\frac{2g_\sigma \omega_c}{E_\sigma^2 - \omega_c^2}\right\} + \mathcal{O}\left(\frac{g_{\tau(\sigma)}^2}{|\omega_c - E_{\tau(\sigma)}|^2}\right). \quad (3)$$

It is important to note that the second order terms that allow photon emission with the same frequency as the zero order terms need to be taken into account since they will add up coherently and the crossed terms contribute to second order to the final master equation. Including such second-order terms, the coupling operator in the dispersive basis reads

$$a + a^\dagger \to (a + a^\dagger)\left[1 + \sum_{i=1,2}\left\{\frac{2g_\tau^2 E_\tau \omega_c}{(E_\tau^2 - \omega_c^2)^2}\tau_z^{(i)} + \frac{2g_\sigma^2 E_\sigma \omega_c}{(E_\sigma^2 - \omega_c^2)^2}\sigma_z^{(i)}\right\}\right] - \sum_{i=1,2}\left\{\tau_x^{(i)}\frac{2g_\tau \omega_c}{E_\tau^2 - \omega_c^2} - \sigma_x^{(i)}\tau_z^{(i)}\frac{2g_\sigma \omega_c}{E_\sigma^2 - \omega_c^2}\right\}. \quad (4)$$

By tracing out now the cavity bath, assumed to be at zero temperature (only transitions to lower energy states are possible), this leads to the Liouvillian

$$\mathcal{L}_\kappa \rho^{\mathrm{d}}_{\mathrm{tot}} = \frac{\kappa}{2} \mathcal{D}\left[a\left\{1 + \frac{2g_\tau^2 E_\tau \omega_c}{(E_\tau^2 - \omega_c^2)^2}(\tau_z^{(1)} + \tau_z^{(2)}) + \frac{2g_\sigma^2 E_\sigma \omega_c}{(E_\sigma^2 - \omega_c^2)^2}(\sigma_z^{(1)} + \sigma_z^{(2)})\right\}\right] \rho^{\mathrm{d}}_{\mathrm{tot}}$$
$$+ \frac{\kappa}{2}\left(\frac{2g_\tau \omega_c}{E_\tau^2 - \omega_c^2}\right)^2 \mathcal{D}\left[\tau_-^{(1)} + \tau_-^{(2)}\right] \rho^{\mathrm{d}}_{\mathrm{tot}} + \frac{\kappa}{2}\left(\frac{2g_\sigma \omega_c}{E_\sigma^2 - \omega_c^2}\right)^2 \mathcal{D}\left[\sigma_-^{(1)} \tau_z^{(1)} + \sigma_-^{(2)} \tau_z^{(2)}\right] \rho^{\mathrm{d}}_{\mathrm{tot}}, \qquad (5)$$

where $\rho^{\mathrm{d}}_{\mathrm{tot}}$ is the partial density matrix including the cavity and the two single-electron DQDs. In the limit of an empty cavity, the master equation for the partial density matrix describing the state of the two DQDs, $\rho^{\mathrm{d}} = \mathrm{Tr}_a\{\rho^{\mathrm{d}}_{\mathrm{tot}}\}$, is given by the two last terms. Finally, to first order in $g_\sigma/\Delta$, with $\Delta = E_\sigma - \omega_c$, the effect of the cavity damping consists essentially of a collective spin decay,

$$\mathcal{L}_\kappa \rho^{\mathrm{d}} \simeq \frac{\kappa}{2} \frac{g_\sigma^2}{\Delta^2} \mathcal{D}\left[\sigma_-^{(1)} \tau_z^{(1)} + \sigma_-^{(2)} \tau_z^{(2)}\right] \rho^{\mathrm{d}}. \qquad (6)$$

Here, $\gamma_\kappa = \kappa \frac{g_\sigma^2}{\Delta^2}$ is the well-known Purcell rate for dispersive decay.

### III. PHONON RELAXATION

The phonons in the host material of each DQD create an electric field which couples to the electronic charge via the electric dipole moment $\tilde{\tau}_z^{(i)}$. To obtain a master equation for the system, we can transform into the system eigenbasis and then trace out the phonon bath, considered here to be at zero temperature. This results in Lindblad terms of the form $\sim \mathcal{D}[\tau_-]$ and $\sim \mathcal{D}[\sigma_-]$, corresponding to charge and spin relaxation respectively, if the charge-cavity coupling is off, $g_c = 0$. If $g_c \neq 0$ we need to transform the electric dipole operator into the dispersive basis before tracing out the phonon bath. This transforms as

$$\tilde{\tau}_z^{(i)} \to -\cos\bar{\phi}\,\tau_x^{(i)} + \sin\bar{\phi}\,\sigma_x^{(i)} \tau_z^{(i)} + 2g_c\left(\frac{E_\tau \cos^2\bar{\phi}}{E_\tau^2 - \omega_c^2}\tau_z^{(i)} + \frac{E_\sigma \sin^2\bar{\phi}}{E_\sigma^2 - \omega_c^2}\sigma_z^{(i)}\right)(a^\dagger + a)$$
$$+ 2g_c \sin\bar{\phi}\cos\bar{\phi}\left(\frac{E_\tau}{E_\tau^2 - \omega_c^2}\sigma_x^{(i)} \tau_x^{(i)} + \frac{E_\sigma}{E_\sigma^2 - \omega_c^2}\sigma_y^{(i)} \tau_y^{(i)}\right)(a^\dagger + a)$$
$$+ 2ig_c \sin\bar{\phi}\cos\bar{\phi}\frac{(E_\tau^2 - E_\sigma^2)\omega_c}{(E_\tau^2 - \omega_c^2)(E_\sigma^2 - \omega_c^2)}\sigma_x^{(i)} \tau_y^{(i)}(a^\dagger - a) + \mathcal{O}(g_c^2). \qquad (7)$$

This means that the phonons will induce $\tau$-conserved and $\sigma$-conserved relaxation (first and second terms), photon induced dephasing (third term), and relaxation to states with different $\sigma$ and different $\tau$ via photon absorption and/or emission.

If we trace out the phonon bath, considered at zero temperature, and in the limit of empty cavity, we get the following Liouvillian

$$\mathcal{L}_\gamma^{(i)} \rho^{\mathrm{d}} = \frac{\gamma_{\mathrm{ph}}^{E_\tau}}{2} \mathcal{D}\left[\cos\bar{\phi}\,\tau_-^{(i)} + \mathcal{O}(g_c^2)\right] \rho^{\mathrm{d}} + \frac{\gamma_{\mathrm{ph}}^{E_\sigma}}{2} \mathcal{D}\left[\sin\bar{\phi}\,\sigma_-^{(i)} \tau_z^{(i)} + \mathcal{O}(g_c^2)\right] \rho^{\mathrm{d}}$$
$$+ 2\left[g_c(E_\tau - E_\sigma)\cos\bar{\phi}\sin\bar{\phi}\right]^2 \frac{\gamma_{\mathrm{ph}}^{E_\tau + E_\sigma - \omega_c}}{(E_\tau - \omega_c)^2 (E_\sigma - \omega_c)^2} \mathcal{D}\left[\sigma_-^{(i)} \tau_-^{(i)}\right] \rho^{\mathrm{d}}$$
$$+ 2\left[g_c(E_\tau + E_\sigma)\cos\bar{\phi}\sin\bar{\phi}\right]^2 \frac{\gamma_{\mathrm{ph}}^{E_\tau - E_\sigma - \omega_c}\Theta(E_\tau - E_\sigma - \omega_c)}{(E_\tau - \omega_c)^2 (E_\sigma + \omega_c)^2} \mathcal{D}\left[\sigma_+^{(i)} \tau_-^{(i)}\right] \rho^{\mathrm{d}}, \qquad (8)$$

where $\gamma_{\mathrm{ph}}^E$ is the phonon relaxation rate evaluated at the energy $E$. It is important to note that the second order terms, omitted here, can contribute to second order via the crossed terms from the first and second Lindblad operators.

To obtain the important terms in the operating regime $\omega_c \sim E_\sigma$, we consider only first order terms in $g_\sigma/\Delta$

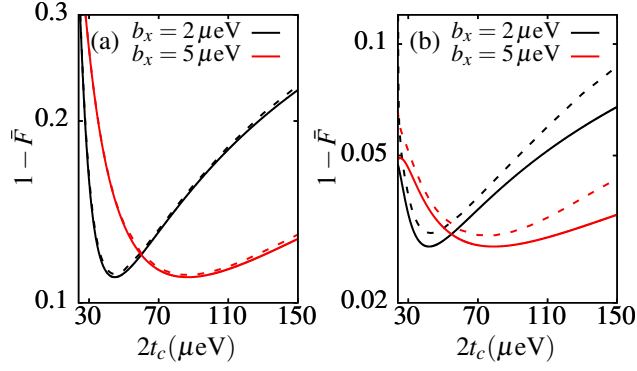

Figure 1. Average infidelity $1 - \bar{F}$ of an iSWAP gate between spin qubits as a function of the tunnel splitting $2t_c$ and in the dispersive regime, $\Delta = -10\, g_\sigma$, for the indicated values of $b_x$ and fixed cavity frequency $\omega_c = 24\,\mu\text{eV}$. The phonon relaxation rate is assumed to be $\gamma_{\text{ph}}^{\omega_c}/2\pi = 4\,\text{MHz}$, the cavity loss rate $\kappa/2\pi = 1.5\,\text{MHz}$, and the charge-photon coupling is chosen as (a) $g_c/2\pi = 50\,\text{MHz}$ and (b) $g_c/2\pi = 200\,\text{MHz}$. The continuous line corresponds to Eq. (6) in the main text, while the dashed lines correspond to the full master equation, Eq. (10).

$(\Delta = E_\sigma - \omega_c)$, obtaining the simplified Liouvillian

$$\mathcal{L}_\gamma^{(i)} \rho^d \simeq \frac{\gamma_{\text{ph}}^{E_\tau}}{2} \cos^2 \bar{\phi}\, \mathcal{D}\left[\left(1 - \frac{g_\sigma^2}{\Delta^2}\right)\tau_-^{(i)} - \frac{g_\sigma^2}{\Delta^2}\left[\sigma_z^{(i)} + \left(\sigma_+^{(i)}\sigma_-^{(j\neq i)} - \sigma_-^{(i)}\sigma_+^{(j\neq i)}\right)\tau_z^{j\neq i}\right]\tau_-^{(i)}\right]\rho^d$$
$$+ \frac{\gamma_{\text{ph}}^{E_\sigma}}{2}\sin^2\bar{\phi}\, \mathcal{D}\left[\left(1 - \frac{g_\sigma^2}{2\Delta^2}\right)\sigma_-^{(i)}\tau_z^{(i)} + \frac{g_\sigma^2}{2\Delta^2}\sigma_z^{(i)}\sigma_-^{(j\neq i)}\tau_z^{(j\neq i)}\right]\rho^d$$
$$+ 2\cos^2\bar{\phi}\, \frac{\gamma_{\text{ph}}^{E_\tau+E_\sigma-\omega_c} g_\sigma^2}{\Delta^2}\mathcal{D}\left[\sigma_-^{(i)}\tau_-^{(i)}\right]\rho^d\ . \qquad (9)$$

The crossed terms from the second line scale as $\frac{\gamma_{\text{ph}}^{E_\sigma}\sin^2\bar{\phi}}{4}\frac{g_\sigma^2}{\Delta^2}$, while the ones resulting from the cavity damping, as shown in Sec. II, scale as $\frac{\kappa}{2}\frac{g_\sigma^2}{\Delta^2}$, which means that the former ones are not relevant while $\gamma_{\text{ph}}^{E_\sigma}\sin^2\bar{\phi} \ll 2\kappa$. Finally, we can adiabatically eliminate the higher energy charge sector and write the electron-phonon induced spin qubit relaxation as in Eq. (5) in the main text.

## IV. AVERAGE FIDELITY

In the following we compare the fidelity results obtained for the approximated master equation in Eq. (5) in the main text with the performance using the full master equation

$$\dot{\rho}^d = -i\left[H^d, \rho^d\right] + \mathcal{L}_\kappa \rho^d + \sum_{i=1,2} \mathcal{L}_\gamma^{(i)}\rho^d\ . \qquad (10)$$

As shown in Fig. 1, the difference increases with the coupling strength $g_c$. Here we have assumed that the electron-phonon relaxation rate is $\gamma_{\text{ph}}^E = \gamma_{\text{ph}}^{\omega_c}(E/\omega_c)^5$ [1].

## V. CHARGE NOISE

The DQD qubits are sensitive to external electrical noise mainly via fluctuations in the onsite energy asymmetry $\epsilon$. We can model such a perturbation with the term $\tilde{V} = \delta \tilde{\tau}_z/2$ added to the Hamiltonian in Eq. (1) of the main text, where $\delta$ is the fluctuating parameter. Transformed into the eigenbasis, this reads

$$V = -\frac{\delta}{2}\cos\bar{\phi}\,\tau_x + \frac{\delta}{2}\sin\bar{\phi}\,\sigma_x\tau_z\ . \qquad (11)$$

Now we do a Schrieffer-Wolff transformation with generator $S$, such that $[H_0, S] = -V$. In the transformed basis the Hamiltonian reads $H_0 + H_\delta$, with

$$H_\delta = \frac{\delta^2}{4} \left\{ \frac{\cos^2 \bar{\phi}}{E_\tau} \tau_z + \frac{\sin^2 \bar{\phi}}{E_\sigma} \sigma_z + \frac{\sin \bar{\phi} \cos \bar{\phi}}{E_\tau} \sigma_x \tau_x + \frac{\sin \bar{\phi} \cos \bar{\phi}}{E_\sigma} \sigma_y \tau_y \right\}. \qquad (12)$$

Within the RWA, only the first two terms contribute.

---



\end{document}